\documentclass[lettersize,comsoc]{IEEEtran}
\usepackage{cite}
\usepackage{algorithm}
\usepackage{amsmath,amssymb,amsfonts}
\usepackage{algorithmic}
\usepackage{graphicx}
\usepackage{textcomp}
\usepackage{epstopdf}
\usepackage{stfloats}
\usepackage{setspace}
\ifCLASSOPTIONcompsoc
\usepackage[caption=false, font=normalsize, labelfont=sf, textfont=sf]{subfig}
\else
\usepackage[caption=false, font=footnotesize]{subfig}
\usepackage{bm}

\usepackage[T1]{fontenc}
\usepackage{cite}
\usepackage{amsmath,amssymb,amsfonts}
\usepackage{algorithmic}
\usepackage{graphicx}
\usepackage{textcomp}
\usepackage{xcolor}
\usepackage{bm}
\usepackage{changepage}
\usepackage{amsmath}
\usepackage{amstext}
\usepackage{stfloats}
\usepackage{url}
\usepackage{multirow}

\usepackage{amsmath}

\begin{document}
\vspace{-20pt} 
\title{Channel Estimation for Pinching-Antenna \\Systems (PASS)}
\author{Jian Xiao, Ji Wang,~\IEEEmembership{Senior Member,~IEEE}, and Yuanwei Liu,~\IEEEmembership{Fellow,~IEEE}
	\thanks{Jian Xiao and Ji Wang are with the Department of Electronics and Information Engineering, College of Physical Science and Technology, Central China Normal University, Wuhan 430079, China (e-mail: jianx@mails.ccnu.edu.cn; jiwang@ccnu.edu.cn). 
	
	Yuanwei Liu is with the Department of Electrical and Electronic Engineering, The University of Hong Kong, Hong Kong (e-mail: yuanwei@hku.hk).
	}}
\maketitle
\vspace{-20pt} 
\begin{abstract}
Pinching antennas (PAs) represent a revolutionary flexible antenna technology that leverages dielectric waveguides and electromagnetic coupling to mitigate large-scale path loss. This letter is the first to explore channel estimation for Pinching-Antenna SyStems (PASS), addressing their uniquely ill-conditioned and underdetermined channel characteristics. In particular, two efficient deep learning-based channel estimators are proposed. 1) \emph{PAMoE:} This estimator incorporates dynamic padding, feature embedding, fusion, and mixture of experts (MoE) modules, which effectively leverage the positional information of PAs and exploit expert diversity. 2) \emph{PAformer:} This Transformer-style estimator employs the self-attention mechanism to predict channel coefficients in a per-antenna manner, which offers more flexibility to adaptively deal with dynamic numbers of PAs in practical deployment. Numerical results demonstrate that 1) the proposed deep learning-based channel estimators outperform conventional methods and exhibit excellent zero-shot learning capabilities, and 2) \emph{PAMoE} delivers higher channel estimation accuracy via MoE specialization, while \emph{PAformer} natively handles an arbitrary number of PAs, trading self-attention complexity for superior scalability.
\end{abstract}


%
\IEEEpeerreviewmaketitle

\section{Introduction}
\IEEEPARstart{A}{dvanced} multiple-input multiple-output (MIMO) is a critical enabling technology for sixth-generation (6G) wireless networks, targeting ultra-high-speed data transmission and seamless connectivity. Among emerging MIMO technologies, flexible antenna systems, e.g., fluid and movable antennas, {offer dynamic channel reconfiguration capabilities by leveraging its positional or rotational adjustability to adapt to the evolving spatial distribution of users \cite{10752873}.} However, they still struggle with addressing large-scale path loss and the need for line-of-sight (LoS) links, both of which are essential for high-quality communication. To overcome these challenges, pinching antennas (PAs) offer a novel approach to creating controlled radiation sites. PAs were initially introduced and developed into a prototype by NTT DOCOMO in 2022 \cite{Fukuda2022}. The fundamental concept of PAs relies on the transfer of electromagnetic (EM) waves from a dielectric waveguide to a nearby dielectric material. Compared to typical movable antennas, PAs can achieve movement ranges spanning thousands to tens of thousands of wavelengths and support a larger number of connected antennas on a waveguide. This technology allows for the flexible positioning of PAs, enabling the establishment of adjustable and reliable LoS transceiver links in Pinching-Antenna SyStems (PASS), thereby effectively mitigating large-scale path loss.

The theoretical exploration of PA technology is still in its early stages, yet it has already garnered significant attention from both academia and industry. The authors of \cite{Ding2024} first provided a comprehensive performance analysis for PASS and proposed a non-orthogonal multiple access (NOMA)-enhanced PASS. Subsequently, related works have explored performance analysis \cite{tyrovolas2025performance}, {physical modeling and beamforming} \cite{Wang2025}, and minimum rate maximization for PASS \cite{10909665}. It should be emphasized that the existing works on PA optimization strategies hinge crucially on precise channel estimation.
However, the channel estimation in PASS has not been investigated in-depth due to new challenges. Firstly, PASS introduces a fundamental challenge in channel estimation due to the highly coupled nature of the in-waveguide channel and wireless propagation channel between PAs and users. Each waveguide is fed by only one radio frequency (RF) chain, while it may carry multiple PAs. This setup necessitates the recovery of high-dimensional information from low-dimensional observations, resulting in an ill-conditioned underdetermined recovery problem. Secondly, the in-waveguide channel is deterministic and completely depends on the positions of PAs in the waveguide. Hence, no stacking of pilot signals can produce additional independent linear equations to solve the high-dimensional wireless propagation channel. Moreover, the dynamic nature of PAs with the plug-and-play capability complicates the channel acquisition. The activation locations of a large number of PAs can be adjusted in a specified region to improve user service. Therefore, the wireless propagation channel presents near-field spatial non-stationarity effects due to the large array aperture. 

{To fill this research gap, this letter presents the first investigation into channel estimation for PASS. First and foremost, we extend classic linear channel estimators by adapting the antenna switching strategy for PASS, which provide fundamental channel estimation benchmarks. Furthermore, inspired by recent advances in deep learning (DL)-enabled channel estimation \cite{10415876,10288572,10315065}, we propose two efficient DL models to improve channel estimation accuracy while reducing pilot overhead.}
The first DL estimator, termed \emph{PAMoE}, is built on the mixture of experts (MoE) architecture, which integrates PA positions and pilot signal features through multi-expert mechanisms to adaptively model the dynamic channel distributions in PASS. To enhance the flexibility and scalability of the channel estimation model for dynamic PA counts, we further propose a Transformer-style estimator, termed \emph{PAformer}, which leverages the self-attention mechanism to predict channel coefficients on a per-antenna basis. Numerical results demonstrate that the proposed DL estimators achieve superior channel estimation accuracy with significantly reduced pilot overhead compared to conventional channel estimators. In particular, both PAMoE and PAformer exhibit superior zero-shot learning capabilities on dynamic PA configurations without retraining operations. 
\section{System Model and Problem Formulation}
\begin{figure}[t]
	\centerline{\includegraphics[width=2.5in]{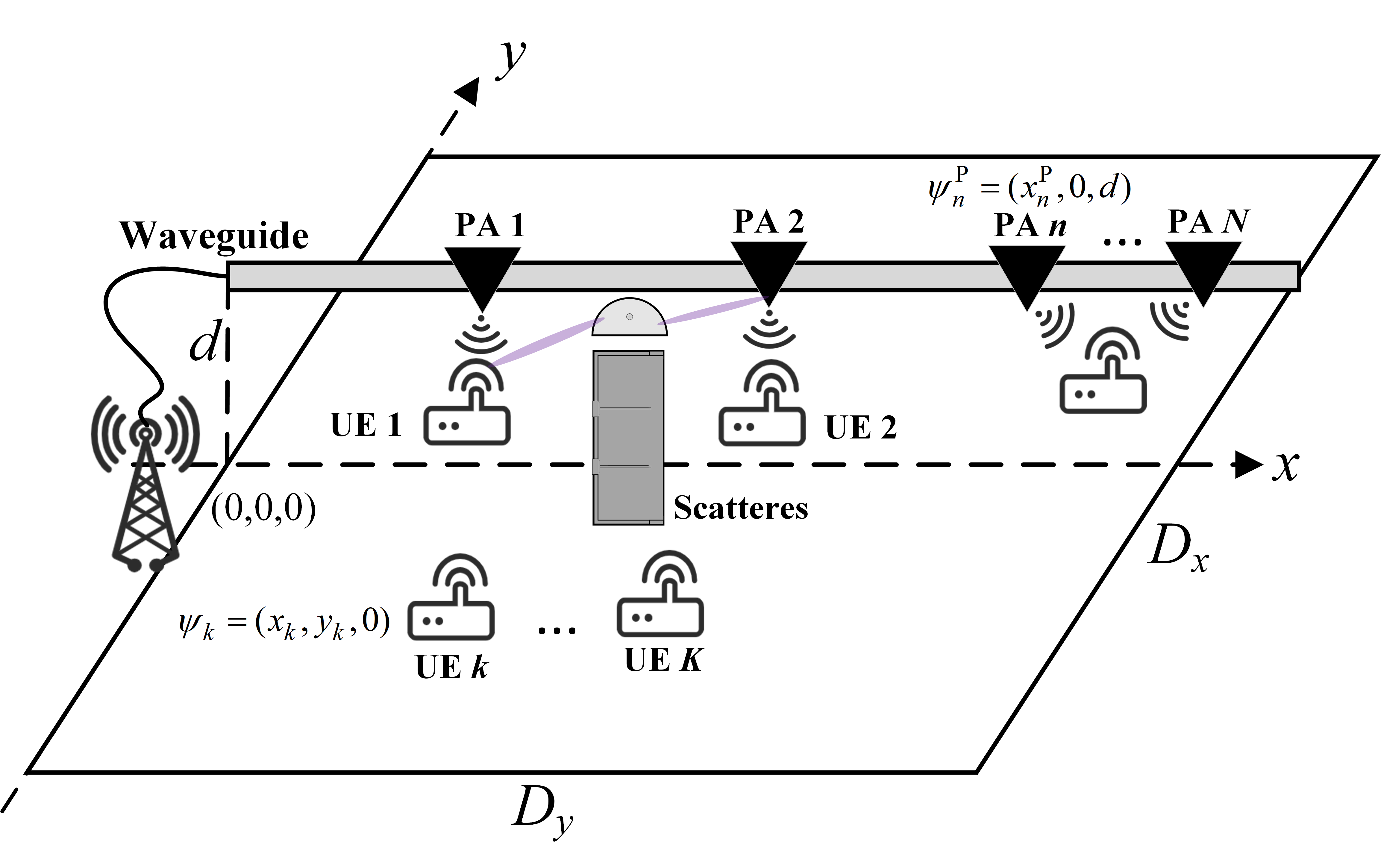}}
	\caption{{Pinching antennas assisted multi-user systems.}}
	\label{fig1}
\end{figure}
As illustrated in Fig.~\ref{fig1}, considering an uplink communication system that includes a base station (BS) equipped with $N$ PAs and $K$ single-antenna user equipments (UEs). Each PA has the same length $L$ on the waveguide. In a Cartesian system, the UEs are assumed to be randomly distributed within a rectangular region on the $x$-$y$ plane, with dimensions $D_x$ and $D_y$. The position of the $k$-th UE is represented by $\psi_k = (x_k, y_k, 0)$. Suppose the waveguide extends parallel to the \(x\)-axis. Its height is denoted by \(d\), and its length aligns with the rectangular dimension \(D_x\). Hence, the coordinates of  PA $n$ are given by \(\psi_n^{\rm{P}} = \bigl(x_n^{\rm{P}}, 0, d\bigr)\), where \(x_n^{\rm{P}}\) lies in the interval \(\bigl[L, D_x\bigr]\). 
In this work, the discrete activation deployment of PAs is adopted, which simplifies the hardware design and is more practical than continuous activation \cite{Wang2025}. The PAs can only be activated at specific discrete positions along the waveguide, forming the feasible set $\mathcal{S} = \left\{ L + \frac{D_x - L}{Q-1}(q-1) \mid q=1,2,\dots,Q \right\}$, where $Q$ denotes the number of discrete positions available.
Since all \(N\) PAs lie along the same waveguide, the transmitted signal of each PA is essentially a phase-shifted version of the signal from the BS at the waveguide feed point. The in-waveguide channel $\mathbf{g}\in {\mathbb{C}^{N\times 1}}$ can be expressed as
\begin{equation}
\begin{split}\label{waveguide}
\mathbf{g} = \begin{bmatrix}
\alpha_1 e^{-j \tfrac{2\pi}{\lambda_g} \left|\psi_0^{\rm{P}} - \psi_1^{\rm{P}} \right|}, \;
\ldots, \;
\alpha_N e^{-j \tfrac{2\pi}{\lambda_g} \left|\psi_0^{\rm{P}} - \psi_N^{\rm{P}} \right|}
\end{bmatrix}^\top,
\end{split}
\end{equation}
where \(\psi_0^{\rm{P}}\) denotes the position of the waveguide’s feed point. Parameter \(\lambda_g = \tfrac{\lambda}{n_e}\) is the guided wavelength. Here, $\lambda$ is the wavelength in free space and \(n_e\) is the effective refractive index of the dielectric waveguide. $\alpha_n$ is the {factor that determines the ratio of power exchanged between the waveguide and PA $n$, governed by the coupling length of the PA} \cite{Wang2025}.

For the wireless propagation channel $\mathbf{h}_{k}\in {\mathbb{C}^{N\times 1}}$ between the $k$-th UE and the PAs, {$\mathbf{h}_{k}$} is composed of the LoS component $\mathbf{h}_{k}^\text{LoS}$ and the non-line-of-sight (NLoS) component $\mathbf{h}_{k}^\text{NLoS}$, i.e., $\mathbf{h}_{k}=\Upsilon_k \odot \mathbf{h}_{k}^\text{LoS}+\mathbf{h}_{k}^\text{NLoS}$. Here, the operator $\odot$ denotes the Hadamard product. $\Upsilon_k=[\upsilon_{k,n},\ldots, \upsilon_{k,N}] \in {\mathbb{C}^{N\times 1}}$ is a Bernoulli random vector and the variable $\upsilon_{k,n}$ takes values from the set $\{0,1\}$, characterizing the existence of a LoS link between the $n$-th PA and the $k$-th UE\footnote{In the existing works for PASS, the free space channel model between UEs and the PAs is commonly utilized by assuming the flexible PAs close to UEs \cite{Ding2024, 10909665,tyrovolas2025performance, Wang2025}. However, in the practical wireless propagation environment, the NLoS channel component caused by the statical and dynamic scatters should be considered. In particular, all PAs are hard to provide the full coverage of LoS services for all UEs in practical deployment environment.}. 
The geometric spherical wavefront model-based LoS channel between the $k$-th UE and the PAs is expressed as \cite{Ding2024}
\begin{equation}
\begin{split}\label{G}
\mathbf{h}_{k}^\text{LoS} = \left[ \frac{\sqrt{\eta} e^{-j \frac{2\pi}{\lambda} \left| \psi_k - \psi_1^{\rm{P}} \right|}}{\left| \psi_k - \psi_1^{\rm{P}} \right|}, \ldots, \frac{\sqrt{\eta} e^{-j \frac{2\pi}{\lambda} \left| \psi_k - \psi_N^{\rm{P}} \right|}}{\left| \psi_k - \psi_N^{\rm{P}} \right|} \right]^\top,
\end{split}
\end{equation}
where \(\eta = \tfrac{\lambda^2}{16\pi^2}\) denotes the path loss at a reference distance of 1 m. 
Considering $S$ scatterers in the $k$-th UE$\to$PAs link, the NLoS channel $\mathbf{h}^\text{NLoS}_k \in {\mathbb{C}^{N \times 1}}$ is given by
\begin{equation}
\begin{split}\label{G}
\mathbf{h}^\text{NLoS}_k = \sqrt{\frac{1}{S}} \sum\limits_{s = 1}^S{\beta_{k,s} \mathbf{a}_{k,s}{e^{j\eta_{k,s}}}},
\end{split}
\end{equation}
where $\beta_{k,s} \sim \mathcal{CN}(0, \sigma_s^2)$ and $\eta_{k,s} \sim \mathcal{U}[0,2\pi]$ denote the complex gain and the random phase of the $s$-th scatterer path, respectively. $\mathbf{a}_{k,s}\in {{\mathbb{C}}^{N\times 1}}$ represents the receiving array response at the PAs. Since the movable region of PAs is large, the equivalent array aperture of PA is likely to exceed the Rayleigh distance that is the criterion to determine the near-field boundary. Hence, the spherical wavefront is utilized to characterize array response $\mathbf{a}_{k,s}$ and is given by 
{\begin{align}
{{{\mathbf{a}}}_{k,s} = \left[ {\frac{\sqrt{\eta}{e^{-j2\pi d_{s,1}/\lambda }}}{d_{k,s}d_{s,1}}, \cdots, \frac{\sqrt{\eta}{e^{-j2\pi d_{s,N}/\lambda }}}{d_{k,s}d_{s,N}}}  \right]^\top},
\end{align}
where $d_{k,s}$ and $d_{s,n},n\in\{1,\ldots, N\}$ denote the distances from UE $k$ to scatterer $s$ and from scatterer $s$ to PA $n$, respectively.}

The received signal at the BS during the $t$-th slot can be expressed as
\begin{equation}
\begin{split}\label{Kchannel}
y_{t} = \sum\limits_{k = 1}^K \mathbf{g}{^\top}\mathbf{h}_{k}s_{k,t} + n_t,
\end{split}
\end{equation}
where \(s_{k,t}\) represents the symbol transmitted by the \(k\)-th UE, and \(n_t \sim \mathcal{CN}(0, \sigma_n^2)\) is the additive white Gaussian noise.

In PASS, the in-waveguide channel $\mathbf{g}$ can be regarded as the deterministic channel component that depends on the locations and the coupling lengths of PAs\footnote{{This work employs the electronic activation of pre-positioned PAs along the waveguide. The system controller only needs to determine the activation state to ascertain the exact positions of PAs. The reconfiguration of the activated PA set is usually several orders of magnitude faster than the channel coherence time, enabling it to respond efficiently to rapid channel variations.}}. Consequently, we {merely need to estimate} the wireless propagation channel $\mathbf{h}_{k}$ from the $k$-th UE to the PAs. However, PASS introduces a fundamental challenge in channel estimation due to the highly coupled nature of $\mathbf{g}$ and $\mathbf{h}_{k}$. Specifically, while each waveguide is connected to multiple PAs, the waveguide channel is fixed and cannot be reconfigured to perform diverse beam measurements as in conventional hybrid precoding. As a result, the system must infer a high-dimensional channel vector $\mathbf{h}_{k}$ from inherently low-dimensional pilot observations, typically just one scalar per waveguide per measurement. To elaborate, suppose the widely used orthogonal pilot transmission strategy, e.g., time-division, is adopted, and $T$ denotes the number of pilot transmission slots transmitted by UE $k$. Note that even if the UE transmits $T \ge N$ or more pilot slots, a single waveguide output may provide no more than one linearly independent measurement per slot. That is, each pilot symbol experiences the same fixed merging of \(N\) PAs in the waveguide, yielding only repeated versions of the same scalar. Hence, the classic error-criterion-based algorithms, e.g., least square (LS) or linear minimum mean square error (LMMSE) estimators, are difficult to apply directly for channel estimation in PASS\footnote{To collect independent pilot measurements for each PA, as required by LS and LMMSE, one feasible approach is to introduce an antenna switching matrix that selects subsets of PAs in different time slots to realize the signal separation. In Section IV of this letter, we provide antenna switching matrix-based LS and LMMSE estimators as channel estimation benchmarks. Note that switching each PA in different slots introduces hardware overhead, extended measurement time, and potential switch losses, all of which reduce practicality.}.

\section{Deep Learning Based Channel Estimation}
In this section, to improve the channel estimation performance, we leverage DL models to develop efficient channel estimation schemes for PASS. However, two critical challenges emerge when applying the DL approach to PASS. Firstly, the channel characteristics vary dramatically as PAs freely change positions within designated regions, which requires the efficient network architecture with sufficient capacity to learn the extended channel state space. {Secondly, by dynamically activating candidate PAs along the waveguide in response to real-time communication demands, the system introduces time-varying channel dimension and heterogeneous signal distribution. The proposed DL-based channel estimator necessitates an innovative architecture capable of adaptively handling spatio-temporal variations in channel parameters, thereby fostering advancements in neural network design to enable robust estimation under non-stationary conditions.}

\subsection{{{Dataset Construction}}}
In the offline training stage, we collect $N_s$ paired samples, i.e., the pilot observation vector $\widetilde{\mathbf{y}}_k {\in {{\mathbb{C}}^{T\times 1}}} $ at the BS, the position set $\Psi^{\rm{P}}=[\psi_n^{\rm{1}},\ldots,\psi_N^{\rm{P}}]=[\bigl(x_1^{\rm{P}}, 0, d\bigr),\ldots, \bigl(x_N^{\rm{P}}, 0, d\bigr)]$ of PAs, and the corresponding channel sample ${\mathbf{h}}_k$. In the proposed channel estimation network, the input tensor is designed as the set of both PA positions and the pilot observation. Considering the waveguide is deployed parallel to the \(x\)-axis, we merely need the \(x\)-axis coordinate set $\mathbf{x}^{\rm{P}}=[x_1^{\rm{P}}, \ldots, x_N^{\rm{P}}]^T {\in {{\mathbb{R}}^{N \times 1}}}$ of PAs as the input feature. {To enable the neural network to adaptively estimate channels across varying PA configurations, we utilize the in-waveguide channel \(\mathbf{g}\) to transform the observation \(\widetilde{\mathbf{y}}_k{\in {{\mathbb{C}}^{T\times 1}}}\) into \(\widetilde{\mathbf{Y}}_k = \mathbf{g}\widetilde{\mathbf{y}}_k^\top \in \mathbb{C}^{N \times T}\). This transformation aligns the input dimension of neural network with the current number of PAs $N$ in PASS, which facilitates the subsequent channel estimation network design.} The complex-value matrix $\widetilde{\mathbf{{Y}}}_k$ is converted into the real-value tensor ${\widebar{\mathbf{Y}}}_k=\{{\Re(\widetilde{\mathbf{{Y}}}_k),\Im(\widetilde{\mathbf{{Y}}}_k)}\}\in{\mathbb{R}}^{N \times 2T}$ for neural network processing. Accordingly, the label tensor in the network training is $\widebar{\mathbf{{H}}}_k=\{{\Re(\mathbf{{h}}_k),\Im(\mathbf{{h}}_k)}\}\in{\mathbb{R}}^{N \times 2}$.

\subsection{PAMoE: Mixture of Experts-Based Channel Estimation}
\begin{figure}[t]
	\centerline{\includegraphics[width=2.5in]{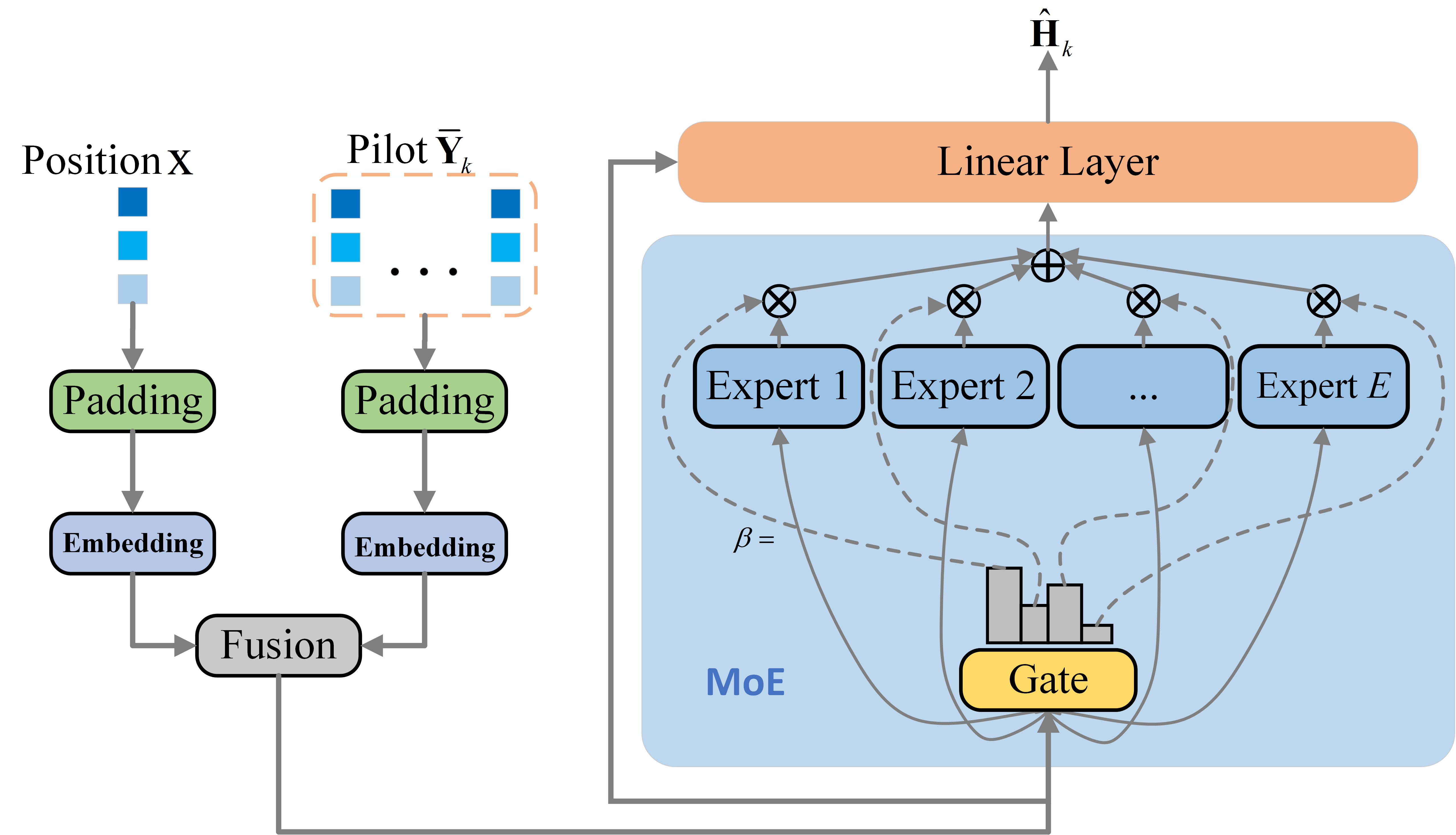}}
	\caption{{Proposed \emph{PAMoE} model for channel estimation.}}
	\label{fig2}
\end{figure}
{We first propose a \emph{PAMoE} estimator as illustrated in Fig.~\ref{fig2} to address high-dimensional dynamic channel estimation from low-dimensional received pilots,} which incorporates the dynamic padding,  feature embedding, fusion, and position-aware MoE modules\cite{cai2024survey}. 

\subsubsection{Dynamic Padding} To accommodate variable PA counts \(N\) and handle variable-length inputs, \emph{PAMoE} employs dynamic padding up to a maximum \(N_{\max}\). Let $B$ denote the batch size in the network training stage. The input tensor in a training batch can be expressed as PA positions \(\mathbf{P} \in \mathbb{R}^{B \times N \times 1}\) and pilot signals \(\mathbf{S} \in \mathbb{R}^{B \times N \times 2T}\), which are the batch version of $\mathbf{x}^{\rm{P}}$ and $\widebar{\mathbf{{Y}}}_k$, respectively. If \(N < N_{\max}\), the network pads along the PA dimension so that both \(\mathbf{P}\) and \(\mathbf{S}\) become length \(N_{\max}\), i.e.,
$\mathbf{P}'= [\mathbf{P}, \mathbf{\phi}^1] \in \mathbb{R}^{B \times N_{\max} \times 1},  
\quad
\mathbf{S}' = [\mathbf{S}, \mathbf{\phi}^2] \in \mathbb{R}^{B \times N_{\max} \times d_{\text{sig}}}$,
where $\mathbf{\phi}^1$ and $\mathbf{\phi}^2 \in \mathbb{R}^{N_{\text{pad}}\times 1} (N_{\text{pad}}=N_{\max}-N)$ denote the learnable padding embeddings, respectively. 

\subsubsection{Feature Embedding} To handle continuous spatial information and achieve efficient extrapolation of PAs, the Fourier basis function {is employed}, developing a Fourier positional embedding approach. Suppose a set of exponentially increasing frequency bases $\mathbf{f} = \left[ 2^f \pi \right]_{f=0}^{F-1} \in \mathbb{R}^{F}$ {is defined,} where \( F \) is the number of frequency components. The scaled position encodings can be expressed as
\begin{equation}\label{sig}
\begin{split}
\mathbf{\Theta} = \mathbf{S}'  \odot \mathbf{f} \in \mathbb{R}^{B \times N_{\max} \times F}.
\end{split}
\end{equation}
Then, we apply sinusoidal transformations to obtain sine and cosine features, and then are concatenated as 
\begin{equation}\label{sig}
\begin{split}
\mathbf{E} = \text{Concat}\bigl[ \sin(\mathbf{\Theta}),\cos(\mathbf{\Theta})\bigl] \in \mathbb{R}^{B \times N_{\max} \times 2F}.
\end{split}
\end{equation}
Further, $\mathbf{E}$ is projected into the embedding space $\mathbf{Z}_{\text{pos}} = \mathbf{E} \mathbf{W}_{\text{pos}} + \mathbf{b}_{\text{pos}} \in \mathbb{R}^{B \times N_{\max} \times d_{\text{embed}}}$,  where \( \mathbf{W}_{\text{pos}} \in \mathbb{R}^{2F \times d_{\text{embed}}} \) is the learnable weight matrix and \( \mathbf{b}_{\text{pos}} \in \mathbb{R}^{d_{\text{embed}}} \) is the bias term. This Fourier embedding effectively captures multi-scale positional variations in the input space.
The pilot signal data is embedded to the same hidden dimension \(d_{\text{hid}}\) by the multilayer perceptron (MLP)-based linear mapping module \(\phi_{\text{sig}}\), i.e., $\mathbf{Z}_{\text{sig}}
= \phi_{\text{sig}} \!\bigl(\mathbf{S}'\bigr)
\;\in\; \mathbb{R}^{B \times N_{\max} \times d_{\text{hid}}}.$
\subsubsection{Feature Fusion} To fuse positional features with pilot signal features, we employ a gating function to generate a gate from a control feature based on PA positions and apply it to the target feature based on pilot signal. The gating operation is formulated as
\begin{equation}\label{sig}
\begin{split}
\mathbf{G}
= \sigma\bigl(\mathbf{W}_g \,\mathbf{Z}_{\text{pos}} + \mathbf{b}_g\bigr)
\;\in\;
\mathbb{R}^{B \times N_{\max} \times d_{\text{hid}}},
\end{split}
\end{equation}
\begin{equation}\label{sig}
\begin{split}
\mathbf{Z}_{\text{fused}}
= \mathbf{G} \,\odot\, \mathbf{Z}_{\text{sig}} + \mathbf{Z}_{\text{pos}}
\;\in\;
\mathbb{R}^{B \times N_{\max} \times d_{\text{hid}}},
\end{split}
\end{equation}
where $\mathbf{W}_g$ and $\mathbf{b}_g$ denote the weight and bias of a linear layer with $d_{\text{hid}}$ neurons, respectively, and \(\sigma(\cdot)\) is a Sigmoid activation function.
\subsubsection{MoE With Gating Network} 
Suppose there are \(E\) experts and each expert takes \(\mathbf{Z}_{\text{fused}}\) and processes it with an MLP-Mixer block across both the feature dimension \(d_{\text{hid}}\) and the spatial dimension \(N_{\max}\). Let
$\mathbf{Z}_e
=E_{e}\!\bigl(\mathbf{Z}_{\text{fused}}\bigr)\in \mathbb{R}^{B \times N_{\max} \times d_{\text{hid}}}$ denote the output of expert $e,(e = 1,\dots,E)$. A gating network is employed to weight each expert’s output. Typically, it pools $\mathbf{Z}_{\text{fused}}$ over the PA dimension to get a global context $\mathbf{z}_{\text{pool}}\in \mathbb{R}^{B \times d_{\text{hid}}}$, and then applies the softmax activation function to produce gating weights, which can be expressed as 
\begin{equation}\label{sig}
\begin{split}
   \boldsymbol{\alpha}
   =
   \mathrm{softmax}
   \bigl(
   \mathbf{W}_\alpha\,\mathbf{z}_{\text{pool}} + \mathbf{b}_\alpha
   \bigr)
   \;\in\;
   \mathbb{R}^{B \times E},
\end{split}
\end{equation}
where $\sum_{e=1}^{E} \boldsymbol{\alpha}_{b,e} = 1, \ \forall b \in \{1,\ldots, B\}$, $\mathbf{W}_\alpha$ and $\mathbf{b}_\alpha$ are the weight and bias of a linear layer with $E$ neurons, respectively. 

Given the output \(\mathbf{Z}_e\) of each expert and the gating weights \(\boldsymbol{\alpha}\), we form a weighted sum over experts. Let \(\boldsymbol{\alpha}\) be reshaped to \(\mathbb{R}^{B \times 1 \times E}\) so it can broadcast over the \(N_{\max}\) dimension to obtain the following output of MoE
\begin{equation}\label{sig}
\begin{split}
\mathbf{Z}_{\text{MoE}}
=
\sum_{e=1}^{E}
\boldsymbol{\alpha}_{e} \,\mathbf{Z}_{e}
\;\in\;
\mathbb{R}^{B \times N_{\max} \times d_{\text{hid}}}.
\end{split}
\end{equation}


Finally, the network concatenates the positional feature \(\mathbf{Z}_{\text{pos}}\) and \(\mathbf{Z}_{\text{MoE}}\) along the last dimension, i.e.,
$
\mathbf{Z}_{\text{concat}}\in
\mathbb{R}^{B \times N_{\max} \times (2\,d_{\text{hid}})}.
$
Then, a linear layer  maps \(\mathbf{Z}_{\text{concat}}\) to the estimated channel $\widetilde{\mathbf{H}}\in \mathbb{R}^{B \times N \times 2}$. In \emph{PAMoE}, the dynamic padding pattern requires the network to fix a predefined maximum number of PAs $N_{\max}$ in the training stage, which is a limiting factor if PAs exceed the predefined maximum bound $N_{\max}$.
\subsection{PAformer: Transformer-Based Channel Estimation}
\begin{figure}[t]
	\centerline{\includegraphics[width=2.5in]{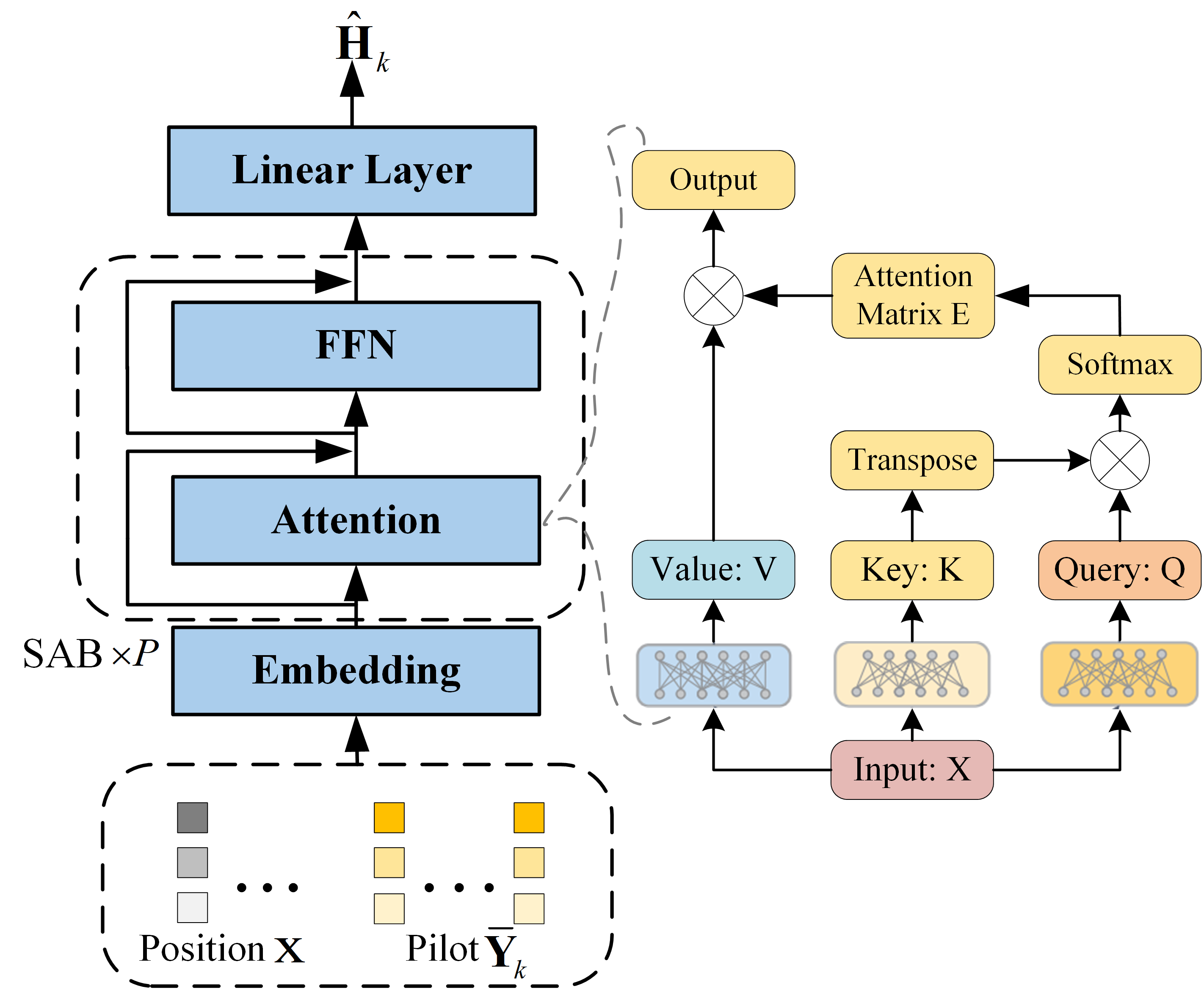}}
	\caption{{Proposed \emph{PAformer} model for channel estimation.}}
	\label{fig3}
\end{figure}
{We further propose a \emph{PAformer} estimator as illustrated in Fig.~\ref{fig3} to address the scalability limitation of \emph{PAMoE}, which is inherently constrained by fixed $N_{\max}$ during training. \emph{PAformer} incorporates self-attention layers with permutation-equivariance, enabling the network to dynamically accommodate PAs of arbitrary size $N$.} The proposed \emph{PAformer} predicts channel coefficients in a per-antenna manner, offering flexibility if new data has more antennas than seen before in the training stage.
\subsubsection{Input Embedding} In \emph{PAformer}, we first concatenate the position and pilot signal features along their last dimension, i.e.,
$\mathbf{V} 
= \bigl[\mathbf{P},\,\mathbf{S}\bigr] 
\;\in\;\mathbb{R}^{B\times N\times(2T+1)}$.
Next, \(\mathbf{V}\) is mapped into a hidden representation \(\mathbf{Z}^{(0)}\) via an embedding network \(\phi\) based on MLP, i.e., $\mathbf{Z}^{(0)} = \phi(\mathbf{V}) 
\;\in\;\mathbb{R}^{B\times N\times d_{\mathrm{hid}}}$.
\subsubsection{Transformer Encoder} \emph{PAformer} employs a pre-norm Transformer with $P$ stacked self-attention blocks (SABs), each comprising multi-head self-attention (MHA) and a feed-forward network (FFN), both with residual connections and layer normalization.  
Let \(\mathbf{Z}^{(\ell-1)}\) denote the input to the \(\ell\)-th SAB, and $\overline{\mathbf{Z}}^{(\ell-1)} = \text{LayerNorm}(\mathbf{Z}^{(\ell-1)})$ is the pre-normalized feature representation by the layer normalization. The output of the MHA module is expressed as $\mathbf{A}^{(\ell)} = \text{MHA} \left(\mathbf{Q}, \mathbf{K}, \mathbf{V} \right) \in \mathbb{R}^{B \times N \times d_{\text{hid}}}$, where $\mathbf{Q}$, $\mathbf{K}$, and $\mathbf{V}$ denote query, key, and value tokens of $\overline{\mathbf{Z}}^{(\ell-1)}$ \cite{vaswani2017attention}, respectively.
The result is combined via a residual connection $ \overline{\mathbf{Z}}'^{(\ell)} = \overline{\mathbf{Z}}^{(\ell-1)} + \mathbf{A}^{(\ell)}$.
Then, the FFN is applied to obtain
$ \mathbf{F}^{(\ell)}
   = \mathrm{FFN}\bigl(\overline{\mathbf{Z}}'^{(\ell)}\bigr),
 $
   where \(\mathrm{FFN}\) is a point-wise MLP operating on each PA token separately. Another residual connection completes this sub-layer
   $\mathbf{Z}^{(\ell)} = \overline{\mathbf{Z}}'^{(\ell)} + \mathbf{F}^{(\ell)}$.
Stacking and applying \(P\) blocks in sequence yields the final feature representation
$\mathbf{Z}^{P} 
= \mathrm{SAB}^{(\ell)}\Bigl(\mathbf{Z}^{(\ell-1)}\Bigr) \in \mathbb{R}^{B\times N\times d_{\mathrm{hid}}},
\ell = 1,\dots,P,$
starting from \(\mathbf{Z}^{(0)}\). 
The final step maps each antenna-wise feature vector in \(\mathbf{Z}^{(P)}\) to the estimated channel $\widehat{\mathbf{H}}_{i}$, where \(i=1,\dots,N\) indexes PAs.
%

For the proposed two DL estimators, during the test stage, the trained network can be applied to scenarios with dynamic numbers of PAs, even though the test dataset has a different data distribution and dimensionality compared to the training dataset. This property of the channel estimation network can also be termed as zero-shot learning.

\section{Numerical Results}
In simulation setups, we set $Q = 200$, $K=4$, $n_e= 1.4$, $d=5$, $D_x \times D_y = 20 \times 20 $ ${\rm{m}}^2$, $S=6$ and the carrier frequency $f_c= 28$ GHz. In the training dataset construction, we collect $N_s= 10^5$ training samples with the fixed number of PAs $N^{\mathrm{tr}} =16$, while the test number of PAs is $N^{\mathrm{te}} \in\{8,\ldots, 32\}$. In the hyper-parameter setups of the proposed DL estimators, we set $N_{\mathrm{max}} = 32$, $E=4$, $d_{\mathrm{hid}}=64$, $P=4$, and $B = 256$. The 1-norm $\ell_1$ is used as the loss function in the network training, i.e., $\ell_1={||\widehat{\mathbf{H}}_k-\widebar{\mathbf{H}}_k||_1}$, while the normalized mean squared error (NMSE) is employed as the performance metric, i.e., $\text{NMSE}=\mathbb{E}\{ {{||\widehat{\mathbf{H}}_k-\widebar{\mathbf{H}}_k||}_{F}^{2}/{||\widebar{\mathbf{H}}_k||_{F}^{2}}}\}$. The antenna switching-based LS and LMMSE estimators are used as the conventional channel estimation benchmarks. {Furthermore, to establish relevant benchmarks incorporating state-of-the-art DL estimators, we further develop two comparison DL benchmarks drawing inspiration from recent attention-based channel estimation networks, i.e., SA-RN-CE \cite{10288572} and DACEN\cite{10315065}, which are created by replacing the MoE module in PAMoE with spatial attention (SA) and dual attention (DA) mechanisms, respectively.} 
\begin{figure}[t]
	\centerline{\includegraphics[width=2.5in]{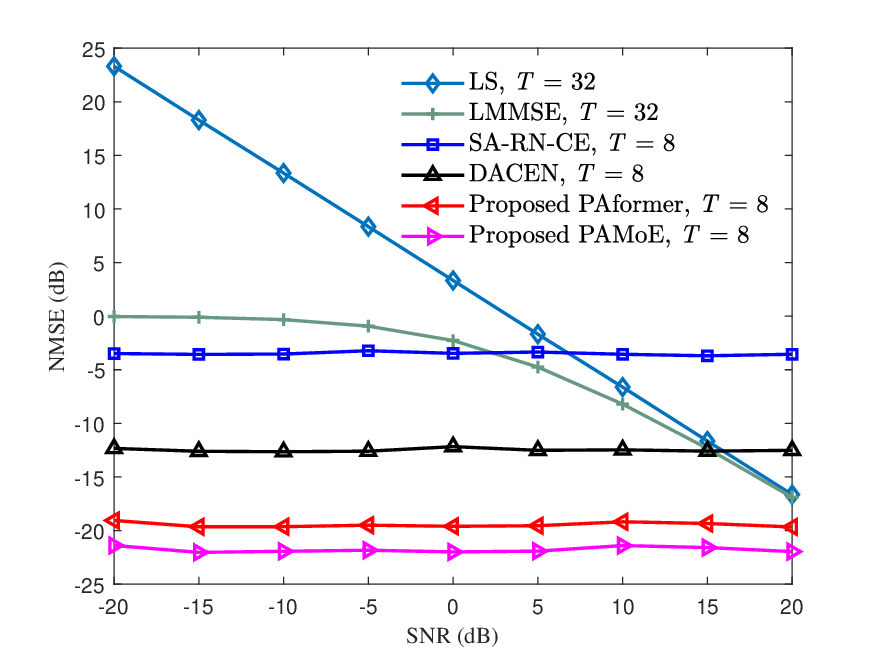}}
	\caption{{NMSE vs. SNR for different algorithms.}}
	\label{fig4}
\end{figure}
\begin{figure}[t]
	\centerline{\includegraphics[width=2.5in]{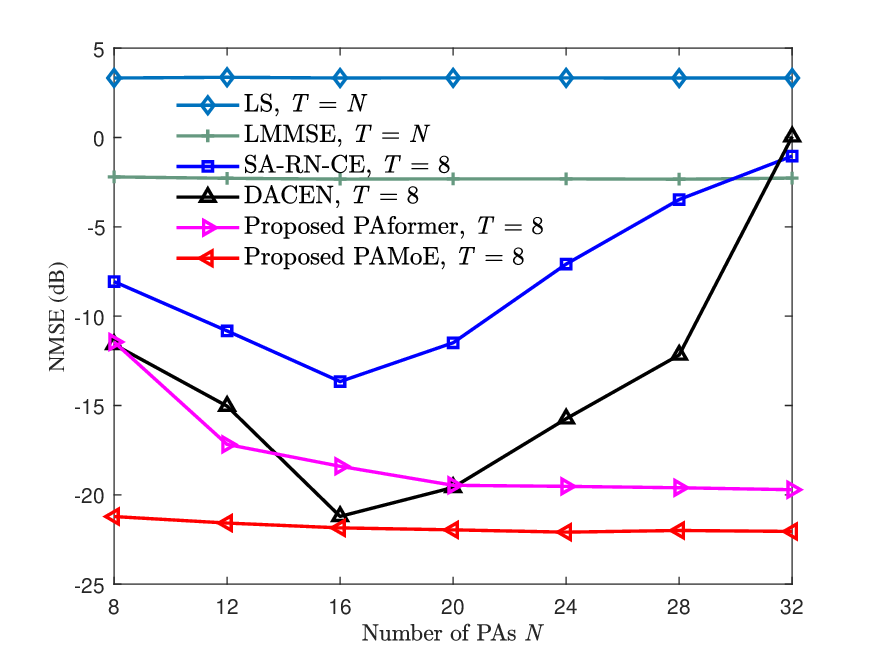}}
	    \captionsetup{labelfont={color=blue}, textfont={color=blue}} 
	 \caption{{NMSE vs. number of PAs $N$ for different algorithms.}}
	\label{fig5}
\end{figure}

In Fig.~\ref{fig4}, we present the NMSE performance of different channel estimation schemes with $N^{\mathrm{te}}=32$. The proposed DL estimators with reduced pilot overhead $T$ outperform the existing linear estimators and DL models. {The superior accuracy of \emph{PAMoE} compared to \emph{PAformer} primarily stems from \emph{PAMoE}'s more specialized design, particularly in its handling and exploitation of the pinching antenna positional information. \emph{PAformer}, while robust and scalable due to its Transformer foundation, operates with a more general mechanism to support variable PA counts.} In Fig.~\ref{fig5}, we provide NMSE performance of different channel estimation schemes for varying numbers of PAs, where the signal-to-noise ratio (SNR) is set to 0 dB. {Compared to SA-RN-CE and DACEN, the proposed \emph{PAMoE} and \emph{PAformer} trained by the fixed PA configuration exhibit excellent robustness and generalization for dynamic PAs.}  The proposed DL estimators possess sufficient zero-shot learning capabilities to deal with distinct data distribution in the test stage. 
\begin{table}[!t] 
    \renewcommand{\arraystretch}{1.3} 
        \captionsetup{labelfont={color=blue}, textfont={color=blue}} 
    \caption{{Complexity Analysis of Proposed DL Estimators}}
    \small 
    \label{tab1} 
    \centering
    {
    \begin{tabular}{cccccc}
        \hline\hline
        \multirow{2}{*}{\textbf{Estimator}} & \multirow{2}{*}{\textbf{$N$}} & \textbf{Params} & \textbf{FLOPs} & \multicolumn{2}{c}{\textbf{Runtime ($\upmu$s)}} \\
        \cline{5-6}
         & & \textbf{(K)} & \textbf{(M)} & \textbf{CPU} & \textbf{GPU} \\
        \hline
        \multirow{3}{*}{\emph{PAMoE}}    & 8  & \multirow{3}{*}{211.9} & \multirow{3}{*}{13.18} & \multirow{3}{*}{211.8} & \multirow{3}{*}{12.73} \\ 
                                  & 16 &                        &     &  &  \\
                                  & 32 &                        &  &  &  \\ 
        \hline
        \multirow{3}{*}{\emph{PAformer}} & 8  & \multirow{3}{*}{545.9} & 4.377  & 74.62 & 3.814 \\ 
                                  & 16 &                        & 8.756     & 128.2  & 7.943  \\
                                  & 32 &                        & 17.51  & 226.8  & 12.95 \\ 
        \hline\hline
    \end{tabular}}
\end{table}
{Table \ref{tab1} summarizes the number of trainable parameters (Params), floating point operations (FLOPs) and inference runtime of the proposed \emph{PAMoE} and \emph{PAformer}, where the NVIDIA RTX 3090 GPU and the 12th Gen Intel(R) Core(TM) i9- 12900K CPU are used as the inference platform. The average inference runtime per channel estimation instance was calculated by averaging over 1000 Monte Carlo experiments. Firstly, for different numbers of PAs $N$, the Params of the proposed \emph{PAMoE} and \emph{PAformer} are the same due to the fixed network architecture. Secondly, the FLOPs of \emph{PAformer} increase linearly with the number of PAs $N$, as the higher-dimensional input tensor needs to be processed in the SAB of \emph{PAformer}. In the \emph{PAMoE} model, the input tensor is uniformly padded to the same $N_{\mathrm{max}}$ via padding interpolation, and hence the FLOPs of the \emph{PAMoE} remain constant for different $N$. Finally, we observe that both \emph{PAMoE} and \emph{PAformer} can achieve the channel estimation at the microseconds ($\upmu$s) level, demonstrating the practical feasibility of the proposed models for real-time deployment.}
\section{Conclusions}

In this letter, we investigated the channel estimation approaches in PASS, and proposed two DL estimators with the advanced neural network architecture to infer a high-dimensional channel vector from inherently low-dimensional pilot observations, respectively. Specifically, the proposed \emph{PAMoE} accommodates variable PA configurations and exploits multi-expert diversity in the MLP-Mixer for improved channel estimation. Accordingly, the proposed \emph{PAformer} is capable of handling arbitrary number of PAs thanks to the self-attention mechanism. Numerical results demonstrated that the proposed DL estimators outperform conventional methods, and significantly reduce the pilot overhead.

\ifCLASSOPTIONcaptionsoff
  \newpage
\fi

\bibliographystyle{IEEEtran}
\bibliography{IEEEabrv,refs.bib}
\end{document}